\pdfoutput=1

\documentclass[11pt,a4paper]{article}
\usepackage[hyperref]{paper_arxiv}
\usepackage{times}
\usepackage{latexsym}

\usepackage{microtype}
\usepackage{inconsolata}

\usepackage{multirow}
\usepackage{graphicx}
\usepackage{url}
\usepackage{amsmath}
\usepackage{amssymb}
\usepackage{cprotect}
\usepackage{enumitem}

\aclfinalcopy 

\title{Text Retrieval with Multi-Stage Re-Ranking Models}

\author{%
	\textbf{Yuichi Sasazawa, Kenichi Yokote, Osamu Imaichi, Yasuhiro Sogawa} \\
	Hitachi, Ltd. Research and Development Group \\ 
	\texttt{\{yuichi.sasazawa.bj, kenichi.yokote.fb,} \\ 
	\texttt{osamu.imaichi.xc, yasuhiro.sogawa.tp\}@hitachi.com}
}

\begin{document}
	
\maketitle
\begin{abstract}
	
	The text retrieval is the task of retrieving similar documents to a search query, and it is important to improve retrieval accuracy while maintaining a certain level of retrieval speed. Existing studies have reported accuracy improvements using language models, but many of these do not take into account the reduction in search speed that comes with increased performance. In this study, we propose three-stage re-ranking model using model ensembles or larger language models to improve search accuracy while minimizing the search delay. We ranked the documents by BM25 and language models, and then re-ranks by a model ensemble or a larger language model for documents with high similarity to the query. In our experiments, we train the MiniLM language model on the MS-MARCO dataset and evaluate it in a zero-shot setting. Our proposed method achieves higher retrieval accuracy while reducing the retrieval speed decay\footnote{The code is available at \url{https://github.com/ckdjrkffz/multi-stage-reranking}.}.

\end{abstract}

\begin{figure*}[t]
	\centering
	\includegraphics[clip, width=16.0cm]{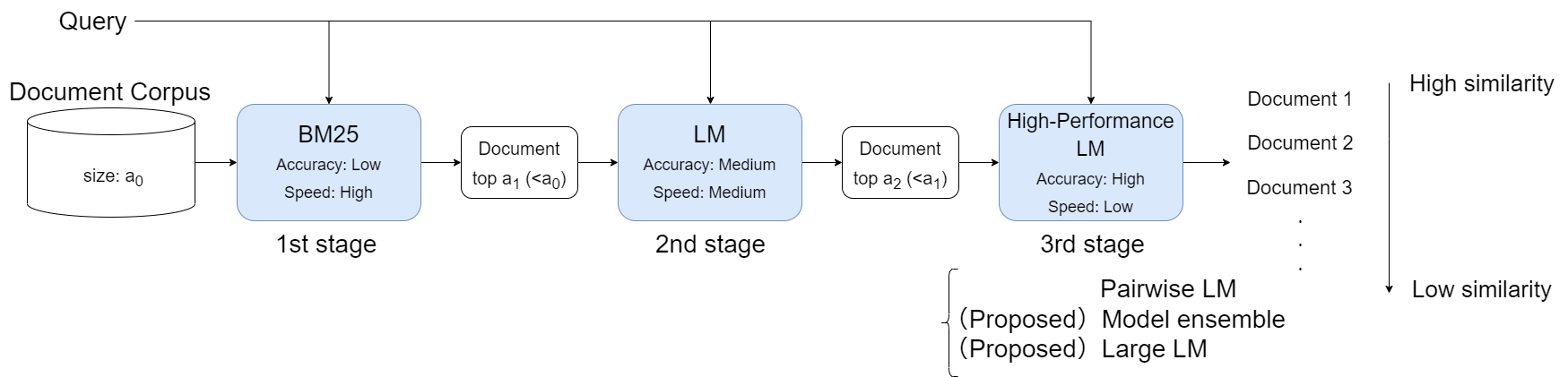}
	\caption{The overview of the multi-stage re-ranking model. In the first stage, the $a_0$ documents are narrowed down to the top $a_1$ documents with high similarity to the query using BM25. In the second stage, these documents are further narrowed down to the top $a_2$ documents with high similarity using the language model. In the third stage, they are further re-ranked by the high-performance model. We compare three types of high-performance models: pairwise language model (existing method), the model ensemble, and the larger language model.}
	\label{fig:overview}
\end{figure*}

\section{Introduction}
The text retrieval is the task of retrieving documents that are similar to a search query, and it is crucial to enhance retrieval accuracy while minimizing retrieval speed delays. Traditionally, lexicon-based methods such as TF-IDF or BM25~\cite{robertson1995okapi} have been used, but recently pre-trained language models such as BERT~\cite{devlin-etal-2019-bert} have been used for text retrieval tasks and have reported high retrieval accuracies. 

One approach using language models is the re-ranking model, which first measures the similarity of each candidate document to the query using BM25. It then re-ranks the documents with the highest similarity by calculating their similarity to the query using the language model. When using the reranking model, similarity calculations using the language model should be applied to the top 100 to 1000 most similar documents. Therefore, using a computationally expensive language model to improve retrieval performance causes a significant delay in retrieval speed. Numerous prior studies have reported accuracy improvements using language models~\cite{https://doi.org/10.48550/arxiv.2202.08904, oguz-etal-2022-domain, nogueira-etal-2020-document, ni-etal-2022-large}, but many of these studies do not considered the trade-off with the reduced search speed associated with increased performance.

To improve search accuracy while reducing search delays, \citet{DBLP:journals/corr/abs-1910-14424} proposed a three-stage re-ranking model. That is, by applying the three models in order as shown below, documents are ranked in order of similarity to the query.

\begin{enumerate}[noitemsep,topsep=0.8pt]
	\setlength{\parskip}{1.0pt}
	\setlength{\itemsep}{1.0pt}
	\item \textbf{BM25} (low accuracy, high speed) calculates the similarity of all documents to the query and ranks them in order of similarity.
	\item \textbf{Language Model} (medium accuracy, medium speed) re-ranks only the documents with the highest similarity (e.g., 100 documents).	
	\item \textbf{High-Performance Language Model} (high accuracy, low speed) re-ranks only the documents with the highest similarity (e.g., 10 documents).
\end{enumerate}

In other words, documents are ranked by BM25 and the language model, and then re-ranked by a high-performance language model for documents with higher similarity. This approach results in a more efficient retrieval process as the highly accurate model is only applied to a limited number of documents.

The high-performance model employed by \citet{DBLP:journals/corr/abs-1910-14424} is a pairwise language model. The pairwise model is a model that classifies which of two documents is more similar to the query, and it is known to be better suited for the document ranking task than the pointwise model, which classifies the correspondence between a query and a single document~\cite{INR-016}. However, the following problems exist with the three-stage re-ranking model using a pairwise model. (1) The pairwise model performs inference on all pairs of documents with high similarity, so the number of inferences is proportional to the square of the number of documents, and the computational cost is very high. For example, ranking the top 10 similarity documents requires $10 \times (10-1) = 90$ inferences. (2) The pairwise model receives a concatenation of a query and two documents. As a result, the sequence of tokens becomes long and often exceeds the input length limit of a general language model. (3) As shown in the experimental results of Section~\ref{sec:exp_result}, the pairwise model performs very poorly in the zero-shot setting for out-of-domain datasets.

In this paper, we propose a three-stage model that uses a model ensemble or a larger model (i.e, models with more parameters than the second-stage language model) at the second stage, as the high-performance model instead of a pairwise model. Model ensembles and models with a large number of parameters have been reported to perform well on a wide range of natural language processing tasks. They are usually unsuitable for text retrieval, where retrieval speed is important, due to increased computational complexity. However, using these models for a limited set of highly similar documents can improve accuracy while minimizing retrieval time. 

In our experiments, we train the language model MiniLM~\cite{wang-etal-2021-minilmv2} on the MS-MARCO dataset~\cite{nguyen2016ms} and evaluate it in zero-shot setting using BEIR datasets~\cite{thakur2021beir}. The reason for evaluating in the zero-shot setting is that many search systems are applied in the zero-shot setting because it is costly to create large training datasets for each task in text retrieval. Experimental results confirm that the proposed method achieves higher retrieval accuracy while reducing the retrieval time compared to existing methods.

\section{Method}
The text retrieval task receives a query as input and ranks documents in the document corpus in order of similarity to the query. The document corpus is common to the training and test sets.

The overview of multi-stage model used in this study in Figure~\ref{fig:overview}. Multi-stage model consists of $n$ modules, and the $i$th module ($0 \leq i \leq n-1$) receives a search query and $a_i$ documents. Then, the module sorts the documents in order of similarity to the query, and outputs the $a_{i+1}$ documents with the highest similarity. $a_0$ is the total number of documents in the corpus.

The advantage of the multi-stage model is that it is easy to make a trade-off between search accuracy and search speed by adjusting the number of documents $a_i$ received by each module. In other words, it is possible to increase $a_i$ if one wants to increase the search accuracy, and to decrease $a_i$ if one wants to increase search speed. The trade-off between accuracy and speed can also be adjusted by increasing or decreasing the number of modules $n$. 

Most existing re-ranking models are two-stage models (i.e., $n=2$), with BM25 as the first module and a language model as the second module. In this study uses an three-stage model (i.e., $n=3$), with a high performance model as the third module. Details of each module are shown below.

\begin{figure}[t]
	\centering
	\includegraphics[clip, height=3.5cm]{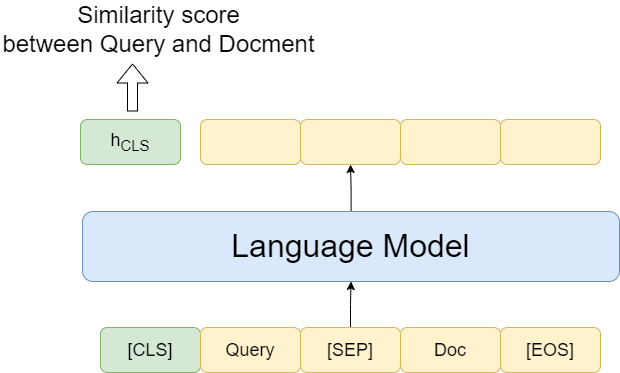}
	\caption{The overview of normal (point-wise) language model. A query and each document are given, and the hidden state corresponding to $\text{\tt [CLS]}$ is given to the classification layer to output the similarity score.}
	\label{fig:model}
\end{figure}

\begin{figure}[t]
	\centering
	\includegraphics[clip, height=3.5cm]{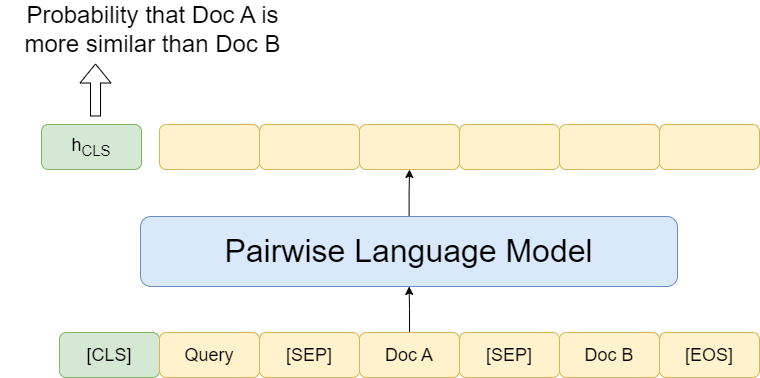}
	\caption{The overview of a pairwise language model. A query and two documents are given, and the hidden state corresponding to $\text{\tt [CLS]}$ is given to the classification layer to output the score for which document is more similar to the query.}
	\label{fig:model_pairwise}
\end{figure}

\subsection{BM25} 
BM25~\cite{robertson1995okapi} is a fast vocabulary-based search method, used by many search systems such as ElasticSearch. BM25 calculates the similarity to the query for all documents.

\subsection{Language Model} ~\label{sec:method_lm}
We calculate the similarity between the query and each document using a cross-encoder language model. That is, as shown in Figure~\ref{fig:model}, the model is given a concatenation of the query and the document, and output the hidden state of each token. Then, the hidden state of $\text{\tt [CLS]}$ token is given to the classification layer. During training, we train the model in the form of binary classification using cross-entropy loss. During inference, we use the scalar value output by the classification layer as the similarity between the query and the document.

\subsection{High-Performance Model}
The top documents with high similarity to the query computed by the language model are re-ranked using a higher-performance similarity calculation method. We compare model ensembles, a larger model, and pairwise models as high-performance models.

\paragraph{Model Ensemble}
We calculate the similarity using an ensemble of language models. That is, multiple models of the same form as the normal language models described in Section~\ref{sec:method_lm} are trained, and the average of the similarity between the query and the document output by each model is used as the similarity by model ensemble. In this study, the model type, the hyperparameters, and the training data of each model used in the ensemble are all identical, differing only in the random seed used to initialize the classification layer and shuffle the training data. While it is generally known that generalization performance can be improved in model ensembles by using multiple models and training conditions, this study shows that performance can be improved by changing a minimum number of settings.

\paragraph{Larger Model}
We use a language model with a larger number of parameters than the second stage language model. In this study, a model with a hidden layer size of 384 is used as a normal language model, and a model with a hidden layer size of 768 as a large model. Similarity calculations are performed in the same way as the method described in Section~\ref{sec:method_lm}.

\paragraph{Pairwise Model}
~\citet{DBLP:journals/corr/abs-1910-14424} proposed a three-stage model using a pairwise model, and we follow this method. As shown in Figure ~\ref{fig:model}, the model is given a concatenation of the query and two documents, and outputs the hidden state of each token. Then, the hidden state of $\text{\tt [CLS]}$ token is given to the classification layer. During training, we train the model in the form of binary classification that makes predictions about whether ``Document A'' or ``Document B'' is more similar to the query. These two documents are extracted from the document corpus so that one corresponds to the query and the other is a negative example that does not correspond to the query. During inference, all pairs of documents with the top $a_2$ similarity (i.e., $a_2 \times (a_2-1)$ pairs) are given to the pairwise model. Let $p_{ij}$ be the score output by the model when ``Document A'' is the $i$th document and ``Document B'' is the $j$th document, and $\sum_{j}(p_{ij} + (1-p_{ji}))$ be the similarity of the $i$th document.


\begin{table*}[!t]
	\centering
	\scalebox{1.0}{
		{\tabcolsep=8.0pt
			\begin{tabular}{lccccc} \hline
				Dataset & \#Train query & \#Test query & \#Doc & Avg. Query Len & Avg. Doc Len \\ \hline
				MS-MARCO & 502,939 & 6,980 & 8,841,823 & 7.24 & 76.63 \\
				FiQA-2018 & -- & 648 & 57,638 & 13.49 & 172.42 \\ 
				SciFact & -- & 300 & 5,183 & 19.94 & 303.94 \\ 
				HotpotQA & -- & 7,405 & 5,233,329 & 22.78 & 63.46 \\ \hline
			\end{tabular}
		}
	}
	\caption{Dataset statistics: the number of training queries, the number of test queries, the number of documents (common to the training data and test data), the average number of tokens (subwords) in the query text, and the average number of tokens (subwords) in the document text.}
	\label{tab:data_statistics}
\end{table*}

\begin{table*}[!t]
	\centering
	\scalebox{0.83}{
		{\tabcolsep=5.0pt
			\begin{tabular}{l|c|cccc|c|c} \hline
				\multirow{2}{*}{Method} & \multicolumn{1}{|c|}{In-domain$\uparrow$} & \multicolumn{4}{|c|}{Out-of-domain (zero-shot)$\uparrow$} & Search time$\downarrow$ & Search delay$\downarrow$ \\
				& MS-MARCO & FiQA-2018 & SciFact & HotpotQA & Average & (sec/query) & (vs BM25+LM) \\ \hline
				BM25 							& 0.2294 & 0.2873 & 0.6642 & 0.6787 & 0.5434 & 0.0178 & ×0.01--0.33 \\
				BM25 + LM 						& 0.3714 & 0.3612 & 0.6518 & 0.7154 & 0.5761 & 0.1939 & ×1.00 \\
				BM25 + LM + Pairwise 			& \textbf{0.3889} & 0.3201 & 0.5450 & 0.6664 & 0.5105 & 0.7759 & ×4.00 \\
				BM25 + LM + Ensemble 	        & 0.3761 & 0.3722 & \textbf{0.6678} & 0.7367 & 0.5922 & 0.2910 & ×1.50 \\
				BM25 + LM + Large	            & 0.3845 & \textbf{0.4141} & 0.6545 & \textbf{0.7843} & \textbf{0.6176} & 0.2693 & ×1.39 \\ \hline \hline
				BM25 + Ensemble					& 0.3761 & 0.3729 & 0.6677 & 0.7388 & 0.5932 & 0.5013 & ×2.59 \\
				BM25 + Large					& 0.3827 & 0.4231 & 0.6522 & 0.7978 & 0.6244 & 0.3966 & ×2.05 \\ \hline
			\end{tabular}
		}
	}
	\caption{We report the retrieval accuracy on the in-domain dataset (MS-MARCO) and on the out-of-domain dataset (FiQA-2018, SciFact, HotpotQA). The evaluation metric of retrieval accuracy is NDCG@10. We also show the average search time on the four datasets and how many times longer each method is compared to BM25+LM. Since there is a large difference in the ratio of search times across datasets for BM25, we show the minimum (SciFact) and maximum (HotpotQA) values.}
	\label{tab:result}
\end{table*}

\section{Experiment}

\subsection{Experiment setting} \label{sec:exp_setting}

The training data is MS-MARCO~\cite{nguyen2016ms}. The test data are FiQA2018, SciFact, and HotpotQA from part of the BEIR dataset~\cite{thakur2021beir} and are evaluated in a zero-shot setting. In addition to that, the results of the in-domain evaluation using MS-MARCO as the test data are also reported as reference figures. Statistics for each dataset are shown in Table~\ref{tab:data_statistics}.

We uses the Pyserini~\cite{Lin_etal_SIGIR2021_Pyserini} toolkit to calculate the similarity by BM25. We use MiniLM~\cite{wang-etal-2021-minilmv2} as the language model\footnote{\url{https://github.com/microsoft/unilm/tree/master/minilm}}. MiniLM is a small distilled language model that is commonly used in text retrieval tasks due to its fast inference speed. For the normal language model, each model in the model ensemble, and the pairwise model, we use a 30M parameters model distilled in  RoBERTa-Large~\cite{DBLP:journals/corr/abs-1907-11692} with 6 layers, 384 hidden layer size. For the larger model, we use a 81M parameters model with 6 layers, 768 hidden layer size, and 81M parameters. For the model ensemble, we use three language models. 

Adam~\cite{DBLP:journals/corr/KingmaB14}, with $\beta_1 = 0.9$, $\beta_2 = 0.999$, $\epsilon=10^{-6}$, and L2 regularization factor $0.01$, was used as the optimizer. We use linear warmup over the first 6\% of the training steps and linear decayed. The dropout rate was 0.1 and a batch size was 64. The learning rate was chosen from $\{1 \times 10^{-5}, 2 \times 10^{-5}, 5 \times 10^{-5}\}$ and the number of epochs was chosen from $\{5, 10, 20, 30\}$. Maximum number of input tokens is $512$ for training and $\{256, 512\}$ for inference, depending on the dataset. If the length of the input token sequence was greater than the maximum number, the end of the document was truncated in the non-pairwise model, and two documents were truncated by the same number of tokens in the pairwise model. $a_1$ is fixed at 100, and $a_2$ is set to 20 in the experiment of section~\label{sec:exp_result}. Hyperparameter selection was performed using MS-MARCO test data\footnote{Since the MS-MARCO test data is not publicly available, the evaluation data is used as hyperparameter search and evaluation.}.

When training, negative samples are randomly selected from the $a_1$ documents with the highest similarity obtained by BM25 calcuration for the in-domain inference model, and randomly selected from all documents in the corpus for the out-of-domain inference model\footnote{That is, separate models were used for in-domain and out-of-domain data for the evaluation. We confirmed that this setup improves performance for each dataset.}. We use NDCG@10 for evaluation metric, following existing studies using the BEIR dataset. Therefore, only the top 10 documents in the ranking are referenced for evaluation. The search time is reported as measured using $1 \times \text{V100}$ GPUs. In all experiments, training and inference were performed three times, and the mean score was reported. Inference for the model ensemble was performed only once.

\subsection{Benchmark method}

\paragraph{BM25} We rank the documents by BM25 only.

\paragraph{BM25+LM} First, we retrieve the documents with the top $a_1$ similarity by BM25, then re-rank them using the language model. 

\paragraph{BM25+LM+Pairwise/Ensemble/Large} We retrieve the documents with the top $a_1$ similarity by BM25, then retrieve the documents with the top $a_2$ similarity by the language model, and finally re-rank them with either a pairwise model, the model ensemble, or a larger model.

\paragraph{BM25+Ensemble/Large} We retrieve the documents with the top $a_1$ similarity by BM25 and re-rank them using the model ensemble or a larger model.

\subsection{Result} ~\label{sec:exp_result}
Table~\ref{tab:result} shows the experimental results, showing that the three-stage re-ranking model with the ensemble or larger model (BM25+LM+Ensemble/Large) achieves superior search accuracy compared to BM25 or BM25+LM. In particular, the three-stage model using the larger model achieves extremely high search accuracy compared to the other methods, with the highest average score for the out-of-domain dataset. However, only in the SciFact dataset, the performance is lower than BM25. The three-stage model with model ensemble achieves stable search accuracy, exceeding BM25+LM on all datasets, although the average score is lower than that of the three-stage model with the larger model. In terms of search time, the three-stage model is limited to 1.2 to 1.5 times the delay of  BM25+LM. Compared to BM25+Ensemble/Large, the search time is significantly reduced, and the search accuracy is comparable or slightly less than that of BM25+Ensemble/Large. These results indicate that the three-stage search model can significantly improve search performance with minimal reduction in search speed.

On the other hand, the BM25+LM+Pairwise three-stage model, which is an existing method, achieves high retrieval accuracy when evaluating in in-domain dataset, but its performance significantly drops in the zero-shot setting using out-of-domain datasets. In other words, the pairwise model has poor generalization performance with out-of-domain data, resulting in lower retrieval accuracy than the normal (pointwise) language model. BM25+LM+Pairwise performs particularly poorly on the SciFact dataset. The reason for this may be: the pairwise model receives as input a sequence of tokens concatenating a query and the two documents, which makes the input token sequence approximately double the length of a concatenation of a query and a document. A typical language model such as MiniLM used in this study has a token length limit of 512 tokens, and especially for datasets with long documents such as SciFact (average 303.94 tokens per document), the input token sequence is often truncated at the end of the document due to token limitations. This may lead to the model not receiving enough information, resulting in decreased accuracy. Additionally, the search speed is much slower compared to BM25+LM. This is because the pairwise model conducts inference on all pairs of highly similar documents, so the number of inferences is proportional to the square of the number of documents.

\subsection{Trade-off between accuracy and speed}\label{sec:result_balance}

\begin{figure}[t]
	\centering
	\includegraphics[clip, width=7.5cm]{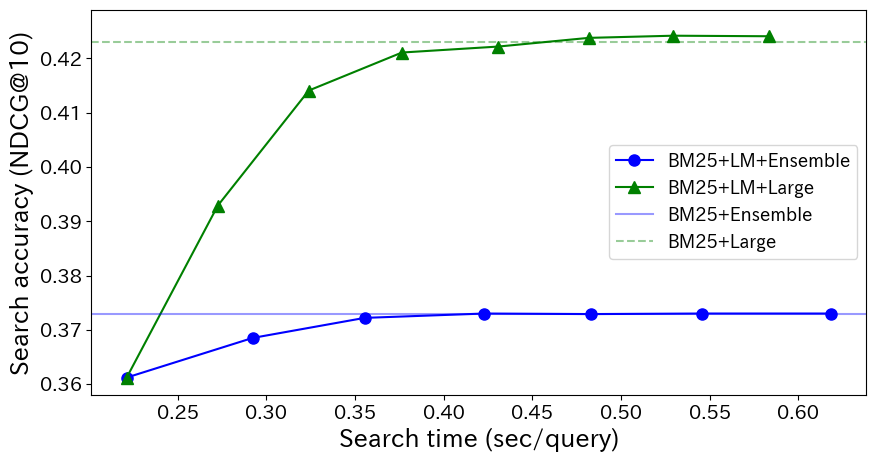}
	\caption{Trade-off between the search accuracy and the search time with $a_2$ varying from 0 to 70. The dataset is FiQA-2018, and the models are BM25+LM+Ensemble and BM25+LM+Large. The search accuracy for BM25+Ensemble and BM25+Large is also shown.}
	\label{fig:result_balance}
\end{figure}

Figure~\ref{fig:result_balance} shows the relationship between search accuracy and search time with $a_2$ varying from 0 to 70.. 

For BM25+LM+Ensemble, the retrieval accuracy converges at $a_2=30$, while for BM25+LM+Large, it converges at $a_2=50$. The accuracy at these convergence points is nearly equivalent to that achieved by BM25+Ensemble and BM25+Large, respectively. In other words, the normal language model has the same ability as the model ensemble  in narrowing down the correct document within 30 candidates, but has a poorer ability to narrow down the candidates to an even narrower range. Therefore, re-ranking the top 30 document candidates retrieved by BM25+LM with a high-performance model is expected to provide higher retrieval accuracy. In fact, the optimal parameters depend on the type of the model and dataset. For example, the larger model used in this study has higher retrieval accuracy than the model ensemble, so the $a_2$ required for convergence of retrieval accuracy is higher than that of the model ensemble. In datasets where document retrieval is straightforward and the gap in accuracy between the normal language model and the high-performance model is minimal, reducing $a_2$ to decrease retrieval time is likely to have a limited impact on accuracy.

\section{Conclusion}

In this study, we proposed a multi-stage re-ranking model using a model ensemble or a larger language model to improve retrieval accuracy while reducing retrieval delay. In our experiments, we trained the MiniLM model on the MS-MARCO dataset and evaluated it in a zero-shot setting. We confirmed that our proposed method achieved higher retrieval accuracy while reducing the retrieval time compared to existing methods. Future work involves verifying the effectiveness of the multi-stage re-ranking model when incorporating alternative modules, such as replacing BM25 with embedding similarity methods like DPR~\cite{karpukhin-etal-2020-dense}.

\bibliography{reference}
\bibliographystyle{acl_natbib}

\end{document}